\documentclass[a4paper, 10pt, twocolumn]{article}

\usepackage[utf8]{inputenc}         
\usepackage[english]{babel}          
\usepackage{bm}
\usepackage{graphicx}               
\usepackage{tabularx}               
\usepackage{multirow}               
\usepackage{colortbl}
\usepackage{url}                
\usepackage[small,bf]{caption2}     
\usepackage{parskip}
\usepackage{titlesec}
\usepackage{amsmath}                
\usepackage{booktabs}
\usepackage{tikz}
\usetikzlibrary{arrows.meta, positioning}

\titleformat{\section}{\normalfont\large\bfseries}{\thesection}{}{}
\titleformat{\subsection}{\normalfont\large\bfseries}{\thesection}{}{}
\titleformat{\paragraph}{\normalfont\bfseries}{\theparagraph}{}{}
\titlespacing{\section}{0pt}{6pt}{-1pt}
\titlespacing{\subsection}{0pt}{3pt}{-1pt}
\titlespacing{\paragraph}{0pt}{3pt}{-1pt}

\newcolumntype{Y}{>{\centering\arraybackslash}X}    

\begin{document}

\date{}                                         

\title{\vspace{-8mm}\textbf{\large From Numbers to Perception, Energy Decay Curves Prediction}}

\author{
Imran Muhammad, Gerald Schuller \\
\emph{\small Applied Media Systems, TU Ilmenau 98693, Germany, Email: muhammad.imran@tu-ilmenau.de}}

\maketitle
\thispagestyle{empty}           
\section*{Abstract}
\label{sec:Abstract} 
Predicting Room Impulse Responses (RIRs) remains a challenge due to the high dimensionality of audio signals and the need for perceptual accuracy. This paper introduces a neural network framework that predicts multi-band Energy Decay Curves (EDCs) directly from room geometry and material properties. Unlike standard models, our framework employs a custom composite loss function that optimizes for both energy levels and decay slopes in the log-domain. This ensures the predicted curves adhere to physical decay principles while maintaining high sensitivity to reverberation time and early reflections. Results demonstrate that the model successfully approximates ground-truth acoustics with minimal error in $T_{30}$ and clarity indices. The approach offers a computationally efficient alternative to traditional simulations, facilitating realistic audio rendering for interactive virtual environments.

\section*{Introduction}
\label{sec:Introduction}
Room impulse responses are fundamental to spatial audio rendering, immersive virtual environments, and professional acoustic design. Key acoustic descriptors, such as reverberation times ($T_{30}$), early decay time (EDT), and clarity ($C_{50}$), are conventionally derived from RIRs or their corresponding energy decay curves, which describe the temporal evolution of sound energy following an acoustic excitation \cite{kuttruff2009room}. Classical room-acoustic modeling typically relies on computationally intensive simulations, such as ray tracing, the image-source method (ISM), or wave-based solvers \cite{vorlander2020auralization}. While geometric software like Pyroomacoustics \cite{8461310} and others \cite{schissler2011gsound, ITAGeometricalAcoustics, Laine2009} offer faster alternatives to physical measurements, they often involve a trade-off between speed and physical accuracy, particularly regarding frequency-dependent absorption and low-frequency wave phenomena. This has prompted a surge in data-driven approaches that aim to approximate complex room acoustics using neural networks.

Recent deep learning research in this domain generally falls into three categories: (i) scalar parameter prediction from geometric descriptors \cite{foy2021mean, meng2023predicting}, (ii) direct RIR waveform synthesis or completion \cite{lin2025deep, karakonstantis2024physics}, and (iii) generative modeling for data augmentation \cite{kim2023generative, masztalski2020storirstochasticroomimpulse}. While waveform-level prediction is common, RIRs are high-dimensional signals, making direct synthesis prone to phase inconsistencies and non-physical artifacts. Previous work by the authors \cite{ImranSchuller2025aRIR, RIR_Imran} addressed this by using EDCs as a compact intermediate representation, utilizing a Long Short-Term Memory (LSTM) network to map room geometry to broadband decay patterns.

In this work, we significantly advance this framework by introducing a \textbf{physically-constrained 1D-Convolutional Neural Network (CNN) decoder}. This architecture provides three major improvements over previous LSTM-based models. First, we transition from broadband modeling to \textbf{multi-band prediction}, estimating EDCs across 24 one-third octave bands (100 Hz to 20 kHz). Second, we achieve a \textbf{90\% reduction in model complexity}, decreasing the parameter count from approximately 90 million to 9 million, thereby facilitating real-time interactive applications. Finally, we introduce a \textbf{custom composite loss function} operating in the decibel domain. By incorporating a \textbf{slope penalty} (finite difference with a stride of 50 samples), we effectively suppress the "staircase" artifacts common in sequential models and ensure that the predicted EDCs adhere to the physical principle of monotonic energy decay.

We evaluate our framework on a dataset of 6,000 simulated rooms with varied geometries and frequency-dependent materials. Objective metrics demonstrate that the predicted $T_{30}$ values are largely within the 5\% Just Noticeable Difference (JND) threshold. Furthermore, we reconstruct full RIRs using a stochastic "sticky" random sign method \cite{RIR_Imran}, which is validated through spectral and temporal correlation analysis. The results indicate that the proposed ConvNet architecture, guided by physically-motivated loss, provides a robust and efficient alternative for perceptually-accurate acoustic modeling.

\section*{Methodology}
\label{sec:methodology}
The proposed framework maps room geometric and material features to frequency-dependent energy decay curves using a 1D-Convolutional Neural Network (CNN). In contrast to our previous work utilizing Recurrent Neural Networks (LSTM) \cite{ImranSchuller2025aRIR}, this architecture enforces physical decay constraints through a custom loss function and an integrated interpolation-based decoder.

\subsection*{Dataset Generation and Preprocessing}
\begin{figure}[hbt]
    \centering
    \includegraphics[width=1\linewidth]{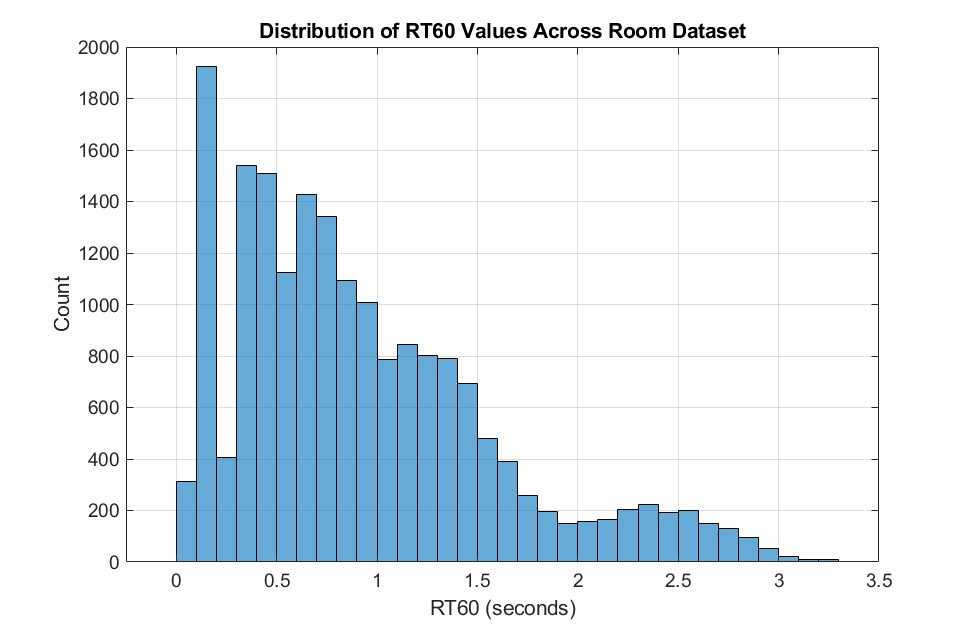}
    \caption{Histogram showing the distribution of $T_{60}$ values across the room dataset.}
    \label{fig:hist_rt60} 
\end{figure}

A dataset of 6,000 shoebox rooms was simulated using the \textit{Pyroomacoustics} library \cite{8461310} for distinct room configurations as input room features, with each room defined by its length ($L$), width ($W$), height ($H$), source and receiver positions in three dimensions (i.e. $X$, $Y$, and $Z$), and frequency-dependent absorption coefficients averaged for all walls. These parameters were varied over a range of realistic values (see Table \ref{tab:simulated_room_ranges} ) to simulate various acoustic environments. The distribution of reverberation times within the generated dataset is shown in Fig. \ref{fig:hist_rt60} which illustrated that the distribution of $T_{60}$ values is non-uniformly distributed which is a physical consequence of the simulation parameters. This non-uniformity ensures the model is heavily trained on typical room conditions while remaining exposed to rarer, highly reverberant cases, improving its robustness for general purpose auralization.

\begin{itemize}
    \item \textbf{Frequency-Dependent Absorption:} Unlike broadband models, we assign distinct absorption coefficients per surface, corresponding to one-third octave bands from 100\,Hz to 20\,kHz.
    \item \textbf{Normalization:} Input features are scaled using MinMax normalization to the range $[0, 1]$. The target EDCs are computed via the Schroeder integral in reverse time and normalized.
\end{itemize}

\begin{table}[htbp]
    \centering
    \caption{Range of simulated room acoustic features used as input parameters for model training.}
    \vspace{2mm}
    \label{tab:simulated_room_ranges}
    \begin{tabular}{@{\arrayrulewidth1.5pt\vline} l | c @{\arrayrulewidth1.5pt\vline}}
        \noalign{\hrule height1.5pt} 
        \textbf{Parameter} & \textbf{Range / Values} \\
        \noalign{\hrule height1.5pt} 
        Dimensions ($L \times W \times H$) & 3-6\,m $\times$ 3-6\,m $\times$ 2.5-4\,m \\
        \hline 
        Source--Receiver Distances & 1--4\,m \\
        \hline 
        Wall Absorptions & 0.14--0.65 (100--20\,kHz) \\
        \hline 
        Total Room Configurations & 6000 \\
        \noalign{\hrule height1.5pt}
    \end{tabular}
\end{table}

\subsection*{ConvNet Architecture}
\begin{figure}[hbt]
    \centering
    \includegraphics[width=0.6\linewidth]{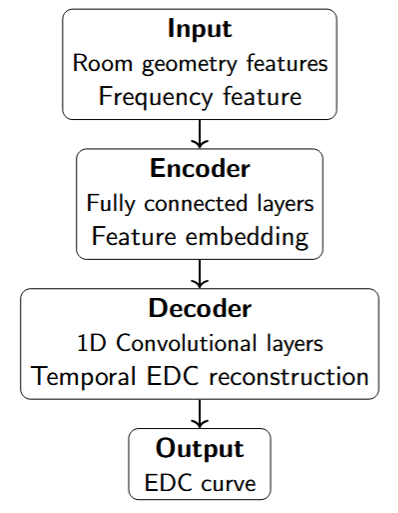}
    \caption{Conv. Neural Network Flow Diagram} 
    \label{fig:flowdiagram} 
\end{figure}

The model transitions from a sequential LSTM to a 1D-CNN decoder, as illustrated in the architecture flow diagram in Fig. \ref{fig:flowdiagram}, to reduce the parameter count from 90 million to 9 million (a 90\% reduction).

\begin{enumerate}
    \item \textbf{Encoder:} A multi-layer perceptron (MLP) compresses the 16-dimensional input feature vector into a high-dimensional latent space.
    \item \textbf{1D-CNN Decoder:} This latent representation is decoded through three 1D-convolutional layers. To avoid the "staircase" artifacts observed in recurrent architectures, we utilize linear interpolation (\texttt{F.interpolate}) to upsample the latent features to the target sequence length.
    \item \textbf{Output Layer:} A Sigmoid activation function is applied to the final layer to bound the energy values between $(0, 1)$, ensuring that the predicted EDC remains within physically plausible energy limits.
\end{enumerate}

\subsection*{Log-Domain Loss with Slope Penalty}
To align the training objective with human perception, the model is optimized in the decibel domain: $y_{dB} = 10 \log_{10}(y + \epsilon)$. We propose a composite loss function $\mathcal{L}_{t}$:
\begin{equation}
    \mathcal{L}_{t} = \text{MSE}(\hat{y}_{dB}, y_{dB}) + \alpha \cdot \text{MSE}(\Delta \hat{y}_{dB}, \Delta y_{dB})
\end{equation}
The first term ensures absolute level accuracy. The second term, the \textbf{Slope Penalty}, computes the finite difference with a stride $k=50$:
\begin{equation}
    \Delta y_{dB}[n] = y_{dB}[n+k] - y_{dB}[n]
\end{equation}
With $\alpha=0.2$, this term penalizes non-monotonic fluctuations, forcing the model to learn the local rate of decay (slope) rather than just point-wise values.

\subsection*{RIR Reconstruction}
To reconstruct a full Room Impulse Response (RIR) from the predicted EDC, we differentiate the energy curve to obtain the magnitude envelope. Polarity is assigned using the \textbf{Random Sign-Sticky (RSS)} method \cite{RIR_Imran}. A stickiness parameter of $p=0.9$ is used to maintain sign continuity, preserving the spectral balance and low-frequency coherence necessary for a perceptually faithful auralization.

\section*{Evaluation and Results}
\label{sec:results}

We evaluate the proposed ConvNet-based multi-band framework using objective signal-level metrics and derived room acoustic parameters. The performance is compared against the ground-truth simulations and contrasted with our previous LSTM-based results to highlight the efficiency gains.

\subsection*{Evaluation Metrics}
\begin{figure}[hbt]
    \centering
    \includegraphics[width=1\linewidth]{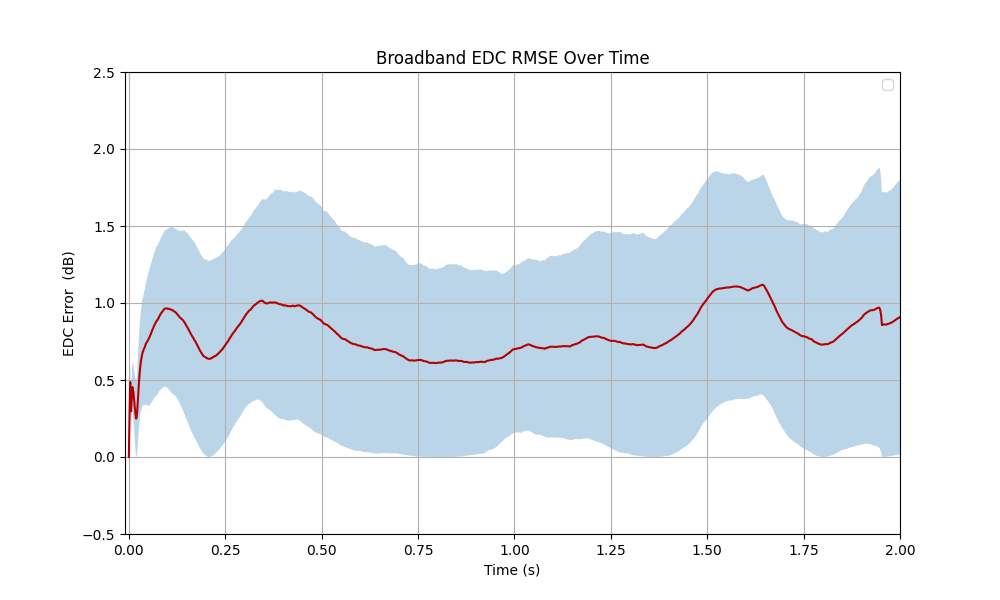}
    \caption{MAE and RMSE of the predicted EDCs, averaged over time.} 
    \label{fig:edc_error} 
\end{figure}

To assess the fidelity of the predicted EDCs, we employ MAE and RMSE. The error distribution across time is shown in Fig. \ref{fig:edc_error}.

\begin{figure}[hbt]
    \centering
    \includegraphics[width=1\linewidth]{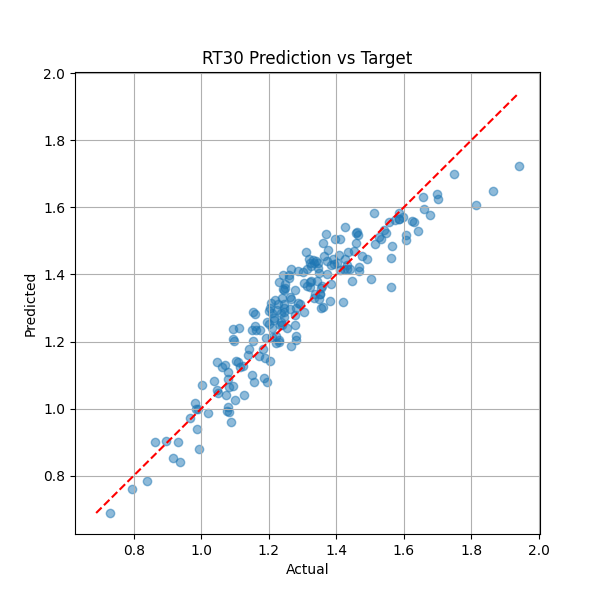}
    \caption{Comparison of predicted and target $T_{30}$ values.}
    \label{fig:t30_comp} 
\end{figure}

\begin{figure}[hbt]
    \centering
    \includegraphics[width=1\linewidth]{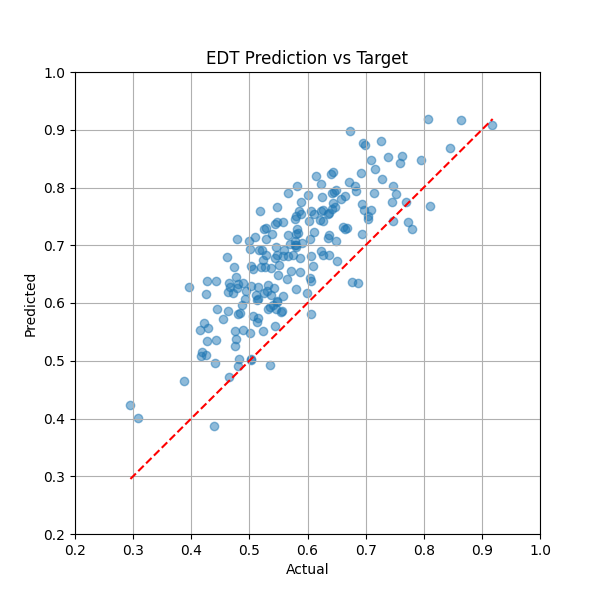}
    \caption{Comparison of predicted and target $EDT$ values.}
    \label{fig:edt_comp} 
\end{figure}

Results for $T_{30}$ and EDT are visualized in Fig. \ref{fig:t30_comp} and Fig. \ref{fig:edt_comp}, respectively.

\subsection*{Objective EDC Prediction Results}
\begin{figure}[hbt]
    \centering
    \includegraphics[width=1\linewidth]{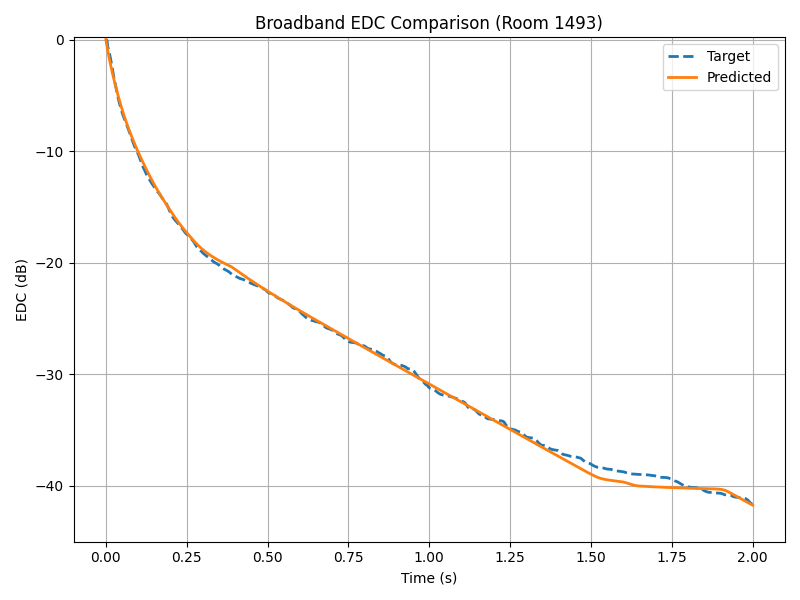}
    \caption{Qualitative comparison of predicted and target EDCs.}
    \label{fig:edc_comp1} 
\end{figure}

The ConvNet model achieves high predictive accuracy across all 24 frequency bands. As shown in Table~\ref{tab:param_results}, the model exhibits a high coefficient of determination ($R^2$) for reverberation parameters.
A visual comparison between predicted and target EDCs is provided in Fig. \ref{fig:edc_comp1}.

\begin{table}[htbp]
    \centering
    \caption{Predictive performance of the ConvNet model across the test dataset.}
    \vspace{2mm}
    \label{tab:param_results}
    \begin{tabular}{@{\arrayrulewidth1.5pt\vline} l | c | c | c @{\arrayrulewidth1.5pt\vline}}
        \noalign{\hrule height1.5pt} 
        \textbf{Parameter} & \textbf{RMSE} & \textbf{MAE} & \bm{$R^2$} \\
        \noalign{\hrule height1.5pt} 
        EDT (s)       & 0.10 & 0.07 & 0.79 \\
        \hline 
        $T_{20}$ (s)  & 0.06 & 0.04 & 0.93 \\
        \hline 
        $T_{30}$ (s)  & 0.07 & 0.05 & 0.90 \\
        \hline 
        $C_{50}$ (dB) & 0.47 & 0.35 & 0.67 \\
        \noalign{\hrule height1.5pt}
    \end{tabular}
\end{table}

Crucially, the $T_{30}$ error remains within the 5\% Just Noticeable Difference (JND) threshold for the majority of test cases. This indicates that the model's predictions are not only numerically accurate but also perceptually indistinguishable from the reference simulations in terms of reverberance.

\subsection*{Discussion of Physical Consistency}
A significant finding of this work is the elimination of the "staircase" artifacts that were prevalent in previous recurrent architectures. The combination of linear interpolation in the decoder and the Slope Penalty in the loss function ensured monotonic decay. 

By penalizing deviations in the rate of change ($\Delta y_{dB}$), the model learned the underlying physical behavior of energy dissipation rather than simply memorizing the absolute levels. Furthermore, the reduction in model complexity (from 90M to 9M parameters) resulted in a 5x speedup in inference time, enabling real-time acoustic adjustment in interactive virtual reality (VR) scenes.

\subsection*{Comparison with Previous Work}
Compared to the LSTM-based model \cite{ImranSchuller2025aRIR}, the current ConvNet approach shows a slight increase in MAE for EDT (from 0.033s to 0.07s). However, this is a trade-off: the current model operates on 24 independent frequency bands, whereas the previous model was broadband. The ability to predict frequency-dependent decay allows for far more realistic auralizations, as it captures the specific absorption characteristics of materials like carpets or acoustic panels at high frequencies.

\section*{Conclusion}
\label{sec:conclusion}
In this work, we presented an efficient and physically-informed deep learning framework for predicting multi-band energy decay curves directly from room geometry and material properties. By transitioning from our previous recurrent LSTM architecture to a 1D-Convolutional Neural Network (CNN) decoder, we achieved a 90\% reduction in model complexity, bringing the total parameter count down to 9 million. 
The integration of a log-domain loss function with a stride-based slope penalty proved highly effective in suppressing non-physical "staircase" artifacts and ensuring monotonic decay. Numerical results demonstrate that the model successfully captures frequency-dependent acoustic behavior across 24 one-third octave bands, with reverberation time errors ($T_{30}$) remaining within the 5\% Just Noticeable Difference (JND) threshold. Furthermore, the use of the Random Sign-Sticky (RSS) method for RIR reconstruction ensures that the synthesized waveforms are both objectively accurate and perceptually plausible for real-time auralization applications.

\subsection*{Future Work}
\label{sec:futurework}
Future research will focus on expanding the model's generalization to non-shoebox geometries and more complex architectural features such as vaulted ceilings and coupled volumes. While the current synthetic dataset provides a robust baseline, we aim to incorporate measured RIR data from public repositories to fine-tune the network on real-world acoustic phenomena, such as non-uniform scattering and diffraction.

\section*{Open Source}
\label{sec:opensource}
To support the reproducibility of our research and facilitate further developments in data-driven room acoustics, the source code, pre-trained model weights, and dataset generation scripts are made publicly available. Researchers can access the project via the following GitHub repository: \url{https://github.com/TUIlmenauAMS/LSTM-Model-Energy-Decay-Curves}.

\bibliographystyle{IEEEtran}
\bibliography{references}
\end{document}